\documentstyle[10pt,phase1]{article}
\begin{document}

\

\vskip5cm

\Large
\centerline{\bf ~~~ DETERMINATION }
\centerline{\bf ~~~  OF }
\centerline{\bf ~~~  THE HUBBLE CONSTANT}

\bigskip
\bigskip

\centerline{~~~Wendy L. Freedman$^1$}
\normalsize 

\bigskip
\vskip3cm
\vskip3cm

\noindent
$^1$ Carnegie Observatories, 813 Santa Barbara St., Pasadena, CA 91101.

\bigskip
\bigskip

\noindent
{\it Proceedings based on a  debate held at  the conference ``Critical
Dialogs in Cosmology'' in Princeton, June 1996.}

\vfill\eject

\Large
\centerline{\bf DETERMINATION OF THE HUBBLE CONSTANT}
\normalsize
\bigskip

\centerline{ Wendy L. Freedman}

\centerline{  Carnegie Observatories}

\medskip





\section { Introduction }

Cosmology   is a   rapidly    maturing  field, and    it  is currently
experiencing a  healthy confrontation between  theory  and experiment.
This   rapid  progress in  many different  areas  of cosmology has not
removed the longstanding interest in measuring many of the fundamental
cosmological parameters: the expansion rate or Hubble constant, H$_0$,
the average  mass density of the Universe,  $\Omega_0$,  the age of  the
oldest objects in  the Galaxy, t$_0$, and the  issue of whether or not
there is a  non-zero  value   for  the  vacuum energy    density,   or
cosmological constant,  $\Lambda$.  Rather,  the increasingly detailed
predictions  of current theory call  further attention to the critical
importance  of accurately  measuring the cosmological parameters which
define the basic model for the dynamical evolution of the Universe.

For instance, accurate knowledge of the Hubble constant is required to
set the time and length scales at the epoch of equality  of the energy
densities of matter and radiation.  In turn, the  scale at the horizon
plays a  role in fixing the  peak in the  perturbation spectrum of the
early universe and  an accurate knowledge  of the Hubble constant will
allow   a  quantitative comparison  of  anisotropies    in  the cosmic
microwave  background  and theories of   the large-scale structure  of
galaxies.  In addition, while a factor  of two uncertainty persists in
the determination of  H$_0$, constraints  on the density of baryons in
the  early   Universe from nucleosynthesis  are   limited to that same
factor of  2 uncertainty.  Coupled with  the current best estimates of
the ages  of  the oldest stars in globular  clusters in our  Galaxy, a
value of  the Hubble constant at  the high end of the  range of values
currently  being  published, would indicate  a  non-zero value for the
cosmological constant, and therefore require new physics not predicted
{\it  a priori}  in  the  current  standard particle-physics-cosmology
model.   It is therefore  imperative  to improve the  accuracy  in the
value   of the  Hubble  constant  and  overcome  the ``factor-of-two''
uncertainty that has persisted in this field for so long.

Primarily  as  a result     of  new  instrumentation  at  ground-based
telescopes, and most recently with the successful refurbishment of the
Hubble Space Telescope (HST), the extragalactic  distance  scale field
has been evolving   at   a rapid pace.    For this reason, during  the
session  on the Hubble  constant, I chose not  to  debate many  of the
details  that  have been  (historically)  central  to the controversy.
Many of the disagreements that I have with Dr.  Tammann are,  in fact,
based  on the  analysis and  interpretation of data   that are rapidly
being  superseded.  To illustrate  the  kinds of   issues involved for
those outside the field, some examples of the areas of  dispute in the
published literature are listed below.
\medskip

$\bullet$   The choice of   methods  for distance  determination.  For
example,  can photographic measurements of   the  angular diameters of
spiral galaxies  give    distances  to   the required   precision   to
distinguish between the  currently  debated values of  H$_0$?  Sandage
(1993) recently concluded   that H$_0$=43$\pm$ 11 km/sec/Mpc  on  this
basis.  However, there is  no  evidence that  the angular diameters of
spiral galaxies are good standard candles; in fact, the  first test of
this method with  a determination of  a Cepheid   distance  to M100, a
spiral galaxy in the  Sandage  sample, yielded a  distance a factor of
almost  2  less than  predicted  by him  on  the basis  of its angular
diameter (Freedman {\it et al.}, 1994).

\medskip

$\bullet$ A dispute over the exact value of the  recession velocity of
the nearest massive cluster,  the Virgo cluster.   This topic could be
debated  ad infinitum, but it  is clear that due  to  the proximity of
this cluster, both its physically-extended nature,  and  an additional
uncertainty due  to its potential  motion with  respect  to the cosmic
microwave background frame, will preclude a  determination of H$_0$ to
better that a precision of  about $\pm$20$\%$.  Few  astronomers would
disagree that the  determination of H$_0$  to higher accuracy requires
an   extragalactic distance scale that  extends  at least an order  of
magnitude more distant than the Virgo cluster, and  a calibration that
is  independent  of the  Virgo  cluster distance.   (Nevertheless, the
distance to  the Virgo cluster can  provide an independent consistency
check to $\pm$20$\%$.)

\medskip

$\bullet$ Large ($\ge$ 25$\%$) scale errors in photographic photometry
that  was  used (almost exclusively)   until the 1980's  and,  in some
cases, continues until the present day ({\it e.g.,} Cepheids: Tammann
\& Sandage 1968;  Sandage 1983;   Sandage and  Carlson 1983;  type  Ia
supernovae: Sandage and Tammann 1993; Sandage {\it et al.} 1994).

\medskip
$\bullet$ Neglect of the effect of dimming due to dust within galaxies
({\it e.g.,} Sandage 1983; Sandage 1988).  Corrections for the effects
of  dust in addition  to  corrections   for errors in the photographic
photometry resulted  in very large (in  some  cases,  40 - $>$100$\%$)
modifications to the   distances to  galaxies  measured with  Cepheids
({\it e.g.,} Freedman \& Madore 1993).

\bigskip
\bigskip

Historically,   measuring accurate  extragalactic   distances has been
enormously difficult; in   retrospect,  the  difficulties  have   been
underestimated and systematic  errors have dominated.  And still,  the
critical remaining  issue  is to  identify and reduce    any remaining
sources of systematic error.  Rather  than delve into and  debate  the
details of the historical  difficulties in measuring  H$_0$, during my
talk I  raised  a number  of general critical  issues that  need to be
addressed  (by practitioners on both  sides  of the ``debate'') before
this problem can be resolved satisfactorily.

\medskip
\medskip
\bigskip
\bigskip
\bigskip
\bigskip

{\bf 1.}  What  is   required  to   measure an  accurate value   of H$_0$?

\medskip
{\bf 2.}  Given  the   wide range of  H$_0$  values quoted in the
current literature, is there any reason to  believe that the situation
has changed very much at all in the  last couple of decades?  From the
perspective   of someone   working  outside    the  field,   with  new
(discrepant)   values for    the Hubble  constant    continually being
published, it is a fair question to ask if {\it any} progress is being
made.

\medskip

{\bf 3.} Is a measurement  of H$_0$ accurate  to 10$\%$  feasible with
current observational tools?

\medskip
\medskip

\par\noindent
These three questions are considered in turn in Sections 2,  3, and 4,
respectively.

\medskip

\section{  What is required to measure an accurate value of H$_0$?}

\medskip

In  principle, the answer   to  this  first question is  very  simple:
measure  the recession  velocities  and  the  distances to galaxies at
sufficiently large distances  where deviations from the smooth  Hubble
expansion are small, and the Hubble constant  follows immediately from
the   slope of the   correlation between velocity  and  distance.   In
practice, however, the  difficulty in measuring  distances to galaxies
has been longstanding, and  unfortunately, the answer to this question
is likely to vary amongst  theorists and observers; moreover, any  two
observers are likely to hold  different opinions about the accuracy of
a given method.  However,  in a very broad  sense, both observers  and
theorists would likely be satisfied with a method that:

\medskip

\vskip 0.2cm
$\bullet$   is based    upon well-understood  physics, 

\vskip 0.2cm
$\bullet$   operates well out into the smooth Hubble flow (velocity-distances 

greater than 10,000 km/sec), 

\vskip 0.2cm
$\bullet$   can be applied to a statistically significant sample of objects and be

 empirically established to have high internal accuracy, and

\vskip 0.2cm
$\bullet$   be demonstrated empirically to be free of systematic errors.

\medskip
\vskip 0.2cm
The above list of criteria applies equally well  to classical distance
indicators  as to  other  physical methods (in   the latter case,  for
example, the Sunyaev Zel'dovich effect or gravitational  lenses). Many
distance indicators have had only an  empirical basis;  however, where
there is an understanding of the physical  mechanism, the residuals in
an underlying correlation can be understood and perhaps corrected.  At
large  distances, the  uncertainties due to   bulk  flows and peculiar
velocities  become  an insignificant  component   of the  total  error
budget; unfortunately very few  methods currently meet  the second and
third criteria.  All  methods require large, statistically significant
samples.   This is not  yet the  case   for the Sunyaev Zel-dovich  or
gravitational  lens methods, for example, where  samples of only a few
or 2 objects, respectively, are currently available.  The  last point,
of course, (ideally) requires that several distance indicators meeting
the first three criteria be available.

At the present  time,   an  ideal distance  indicator or other  method
meeting all  of the above criteria does  not exist, and measurement of
H$_0$ as high  as  1$\%$ accuracy is clearly  a goal for  the  future.
However, this  brings us to questions number  2)  and  3): what is the
current status  of  the field,  and is a   value of H$_0$  accurate to
10$\%$ feasible  with current observational tools?   A brief review of
recent progress  is given in  Section  3). Lastly,  the   Hubble Space
Telescope  Key Project  on the Extragalactic  Distance Scale  has been
designed to measure H$_0$ to 10$\%$  accuracy.  A review  of the goals
of this project will be given, and recent results presented in Section
4).

\section{ Progress Over the Last Decade }

Dramatic  progress has been  made recently in  measuring both absolute
and  relative  distances.     Moreover,  quantitative  comparisons  of
individual indicators allow numerous cross-checks and estimates of the
external,   in  addition  to  internal,    errors.    Before 1980  the
extragalactic distance scale was based almost entirely on photographic
data with large photometric errors.  With CCDs and near-IR arrays more
accurate photometry   has  become   available, with  corrections   for
reddening,  and tests for effects  of  metallicity now being feasible.
Several new, independent methods for measuring relative distances have
also been  developed  and   tested  extensively.  These  issues    are
discussed in more detail in other recent  reviews ({\it e.g.}, see the
proceedings  from the STScI May 1996  Symposium   on the Extragalactic
Distance Scale edited  by Livio \& Donahue 1997;  van  den Bergh 1994;
Jacoby {\it et al.} 1992).

With  the exception  of  a small number   of  independent methods  for
measuring H$_0$    applied  at  large    distances (for  example,  the
Sunyaev-Zel'dovich method   for clusters   or  the gravitational  lens
time-delay method),  most routes to the extragalactic   distance scale
rely on the calibration of  an additional tier  of (secondary) methods
using the Cepheid period-luminosity relation ({\it e.g.,}  the type Ia
supernovae, Tully-Fisher   relation for   spiral galaxies,  or surface
brightness    fluctuations).  In   principle,  the type  II  supernova
expanding atmosphere method  is independent of   the Cepheid  distance
scale, but also may be calibrated by Cepheids as  an external check on
systematics.  Other indicators  (for   example, the  planetary  nebula
luminosity function (PNLF), and tip of the red giant branch (TRGB)) do
not currently operate  beyond the distance  to the  Virgo cluster, and
hence need  to be tied  into other   methods  that can  be applied  at
greater distances where peculiar velocities are a smaller component of
the  overall  expansion  velocity.  Nevertheless,  the PNLF  and  TRGB
methods  provide    an  essential  check    on  the   consistency   of
Cepheid-plus-other-distance methods  in the range   of overlap.  Since
the absolute scale of  most current distance  indicators  is  obtained
using  Cepheids,  it is clearly  imperative  to  eliminate significant
systematic errors in the Cepheid distance scale.

\subsection{ ~Cepheid Distances to Galaxies }

Significant  progress in the  application of Cepheid  variables to the
extragalactic distance   scale has been made  over the past decade  or so.
Many of the  improvements  have   become  possible due  to advances in
detector technology:  in particular,  the  arrival of linear detectors
sensitive  over a broad  range of wavelengths from  the visible to the
near-infrared (see the reviews by Madore \& Freedman 1991; Jacoby {\it
et al.} 1992; Freedman \& Madore 1996).   The discussion below briefly
summarizes that given in Freedman \& Madore (1996).

The areas where the most dramatic improvements have been made include:
\medskip
\par\noindent
1)  Correction for significant (typically 0.5 mag)  scale errors
in the earlier photographic photometry. 
\medskip
\par\noindent
2) Observations of Cepheids beyond the Magellanic Clouds at BVRI
and in some cases, JHK wavelengths, enabling...
\medskip
\par\noindent
3) ... Corrections for interstellar reddening, and 
\medskip
\par\noindent
4) Empirical tests for the effects of metallicity.

\bigskip

During his talk, Dr.   Tammann stressed  the remarkable consistency of
the H$_0$ determinations undertaken by himself and Dr. A. Sandage over
the past 20 years that  yield a value of  H$_0$ = 55 km/sec/Mpc.  This
consistency is truly remarkable.  The interested reader is referred to
a discussion by Freedman \& Madore (1993) of  the changes to the local
Cepheid  distance scale  over the period   from   1974  to  1993.  For
example, for the nearby galaxies M31 and M33, the  published (apparent
blue)   distance  moduli  changed by  1.24    mag (!)  and   0.48 mag,
respectively.  In the case of M81,  the distance was changed twice (by
a factor of almost two) from 3.3 Mpc in 1974,  to 5.8 Mpc in 1984, and
back down  to 3.6 Mpc in  1994.  It is thus  even more remarkable that
despite these enormous (up-and-down) changes  to the zero point of the
Cepheid distance  scale over this   same 20-year period,  the value of
H$_0$ remained at 55.

\medskip

In the subsequent two sections, the effects of reddening and 
metallicity on the Cepheid distance scale are discussed.

\bigskip
\bigskip

\subsubsection{ ~Reddening}

Twenty years ago  photoelectric BVI  photometry  for Magellanic  Cloud
Cepheids had been obtained by a number of authors (see Feast \& Walker
1987;  Madore 1985 for reviews).   However, for  more distant galaxies
where generally  only   {\it  B}-band  photographic    photometry  was
available, corrections were made only  for {\it foreground} reddening,
but not for  reddening of the Cepheids internal  to the  parent galaxy
under study ({\it   e.g.,}  Tammann \&  Sandage 1968;    Sandage 1983,
Sandage and Carlson 1983; however, see Madore 1976).

\begin{figure} 
\plotfiddle{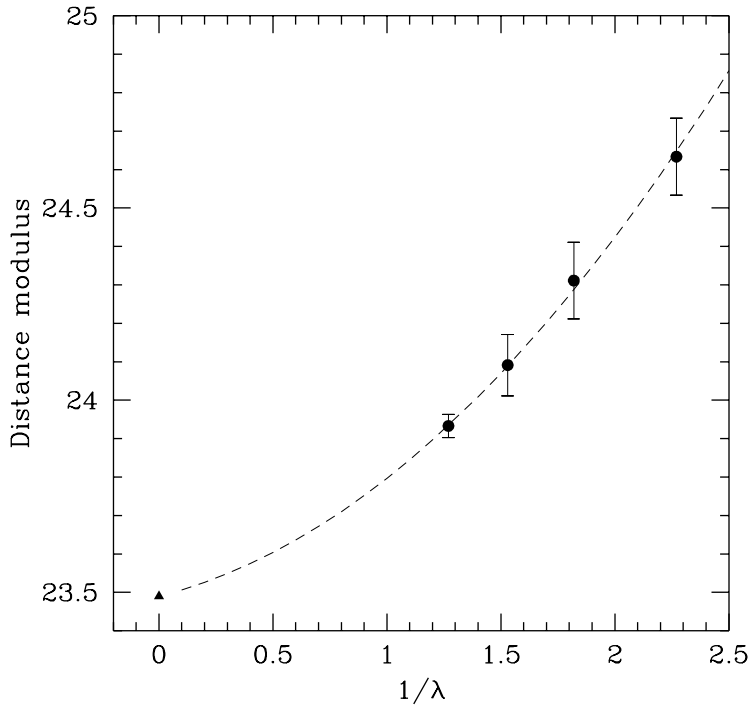}{3.5in}{0}{50}{50}{-160}{-100} 
\medskip
\medskip
\caption{ BVRI apparent  distance  moduli  plotted   as  a function  of
inverse wavelength  for  Cepheids in  NGC~6822.   The  filled triangle
marks the true modulus = intercept of the fit at the origin $1/\lambda
=  0.0$ for E(B--V)  = 0.21 $\pm$ 0.03 mag  (from Gallart  \& Aparicio
1996).  The broken line is a fit of a standard Galactic extinction law
to the data.}
\end{figure} 
\medskip

Recently  it   has become possible  to   determine reddening-corrected
Cepheid  distances to  galaxies based  on  multicolor photometry ({\it
e.g.,  } Freedman 1986; Freedman   1988; Freedman,   Wilson \&  Madore
1990).   This multiwavelength method  has been  adopted  by the Hubble
Space Telescope Key Project Team  on the Extragalactic  Distance Scale
(Freedman {\it et al.}  1994a,b; Ferrarese {\it et  al.} 1995;  Kelson
{\it et  al.} 1995; Silbermann {\it et  al.} 1996; Graham {\it et al.}
1997), and  also by  the other  groups  using HST to  measure  Cepheid
distances  (Sandage {\it et al.} 1994;  Saha  {\it et al.} 1994, 1995,
1996; Tanvir {\it et al.}  1995).  An example  of this multiwavelength
approach  is shown  in  Figure 1 from  a recent  determination of  the
distance modulus to the nearby Local Group galaxy NGC 6822 by Gallart
\& Aparicio (1996).  NGC 6822 sits close to the Galactic plane; hence,
the  foreground  reddening to  this   galaxy is  particularly   large.
However, these data underscore the need  to have multicolor photometry
for determining true moduli, corrected for the effects of interstellar
dust.

\subsubsection{ ~Metallicity}

The   most recent  theoretical  modeling  aimed at  investigating  the
sensitivity of the  Cepheid period-luminosity relation to  metallicity
is that of Chiosi, Wood \& Capitanio (1993).  These authors calculated
linear nonadiabatic  pulsation models for  a   grid of Cepheid masses,
various effective temperatures, and chemical compositions ranging from
1/4 to solar metallicity, for a variety of  mass-luminosity relations.
They conclude that the agreement   between theory and  observation  is
best  at longer wavelengths,  particularly  longward of  the blue band
(where most  historical measurements of Cepheids  were  made).   Their
predicted uncertainty of  the true distance modulus,  after correcting
for  reddening, is $\delta$(m--M)  = --1.7$\delta$Z (at  log P = 1.0).
Hence,  for  the entire range of  chemical compositions represented by
the low-metallicity SMC (Z = 0.004) and the solar-metallicity Galactic
Cepheids (Z = 0.016), a very small abundance effect (amounting to only
0.02 mag, full range) is predicted.

Several observational  programs  are currently  aimed   at undertaking
tests for the sensitivity of the Cepheid period-luminosity relation to
metallicity in the nearby galaxies M31 and  M101.  The prediction of a
small sensitivity to metallicity by  Chiosi  {\it  et al.} (1993) is
consistent with  the  results  of  an observational   test  in M31  by
Freedman \& Madore  (1990).   These  authors undertook a  differential
empirical  test at 3 different fields  located at different radii (and
having different  chemical  compositions) within  M31.   Across  the 3
fields,  they  found a   small   (0.3 mag peak-to-peak)  range in  the
reddening-corrected distance moduli, having a  low overall statistical
significance, but in the same sense  as predicted by theory.  However,
the Freedman \&  Madore data have been  reexamined by Gould (1994) who
comes to the  conclusion that these data  are consistent with a larger
metallicity dependence  than  predicted by   the current models.  This
issue is at present unresolved, and it  is vital  that it be resolved.
It  is  quite independent  of  the other issues  usually  debated over
H$_0$: it potentially could affect all of  the results  based on other
methods   (the   Tully-Fisher    relation,    type   Ia    supernovae,
surface-brightness fluctuations, planetary nebula luminosity function,
type II supernovae,  etc.) all of  which rely for  their calibration on
Cepheid variables.

In this context,   it is interesting  to  note that  there are   other
completely independent limits on the  degree to  which  metallicity is
affecting the Cepheid distance scale.  In particular, and as discussed
in more detail in  the next  section,  distances obtained  using other
methods based on the physics of  stars at completely  different stages
of stellar evolution, and in  some cases ({\it  e.g.,} very metal poor
giant  stars and RR Lyrae  stars) having orders of magnitude different
metallicities, nevertheless yield distances consistent with those from
Cepheids to within  $\pm$10$\%$ rms.  Therefore, both  empirically and
theoretically, there is  no strong  evidence at the present time  that
gross ({\it e.g., }factor of 2) systematic errors exist as a result of
metallicity variations in Cepheid samples.  However, 10-15$\%$ effects
could  well be lurking, and as   the accuracy of distance measurements
continues to  improve, (what  used   to be small!)   effects will  now
dominate the overall error budget.

Fortunately, with the planned  February 1997 instrument  change on the
Hubble Space  telescope,   and  the  subsequent availability  of  the
near-infrared NICMOS camera, there is an  opportunity to vastly reduce
the uncertainty due to metallicity (and reddening) variations by about
a factor of  3.    Bolometrically, the sensitivity to   metallicity is
predicted  to be very  small.   Long  wavelength  ({\it e.g.}, H-band)
observations with NICMOS are currently planned to  address this issue.
In addition,  a 5-year long-term  program  at   Palomar to measure JHK
magnitudes for the M31 Cepheids is now nearing completion.

\subsection{ ~Comparisons with Other Distance Indicators}

As recently as the mid-80's,   some   published Cepheid distances   to
nearby  galaxies were discrepant  by  factors of two  and adopting one
Cepheid calibration over  another could result  in differences  in the
Hubble  constant of almost  a  factor  of two.  Fortunately,  a decade
later, more reliable  distance determinations to  nearby galaxies have
been  obtained  with new  CCD   data using a   number  of  independent
techniques  (Cepheids,  RR Lyraes,  the tip of   the red  giant branch
(TRGB), and even type  II supernovae in the  case  of SN 1987a  in the
LMC).  Moreover, distances to  many  of these same nearby galaxies (as
well as galaxies at intermediate  distances)  have also been  measured
using a number of fairly recently developed  secondary techniques such
as surface brightness fluctuations and the planetary nebula luminosity
function.  Now that photometry with linear  detectors is available for
a variety of  methods, and  corrections for reddening  can be applied,
the distances  to nearby galaxies have  converged  to {\it full range}
differences of less  than 0.3  mag  ({\it i.e.,}  15$\%$ in  distance,
Freedman \&  Madore  1993).   Moreover, the   excellent  agreement  of
individual distances gives no indication of large remaining systematic
errors.

Recent  detailed  discussions   of  the surface-brightness (SBF)   and
planetary-nebula-luminosity-function    (PNLF)   methods are    given,
respectively, by Tonry (1997) and Jacoby (1997).   A comparison of the
Cepheid distances with  those  obtained using both the  SBF   and PNLF
methods yields agreement to   better than  $\pm$ 10$\%$ (1--sigma)  in
distance.  Although  both the SBF and  PNLF distances  are  calibrated
using the  Cepheid  distance to M31,  the  {\it relative} agreement of
these methods  is  extremely  encouraging.  In  Figure  2, the Cepheid
distances     are   plotted   versus  SBF    distances (Tonry, private
communication),  out to  and including  the distance  to   the  Fornax
cluster.

\begin{figure} 
\plotfiddle{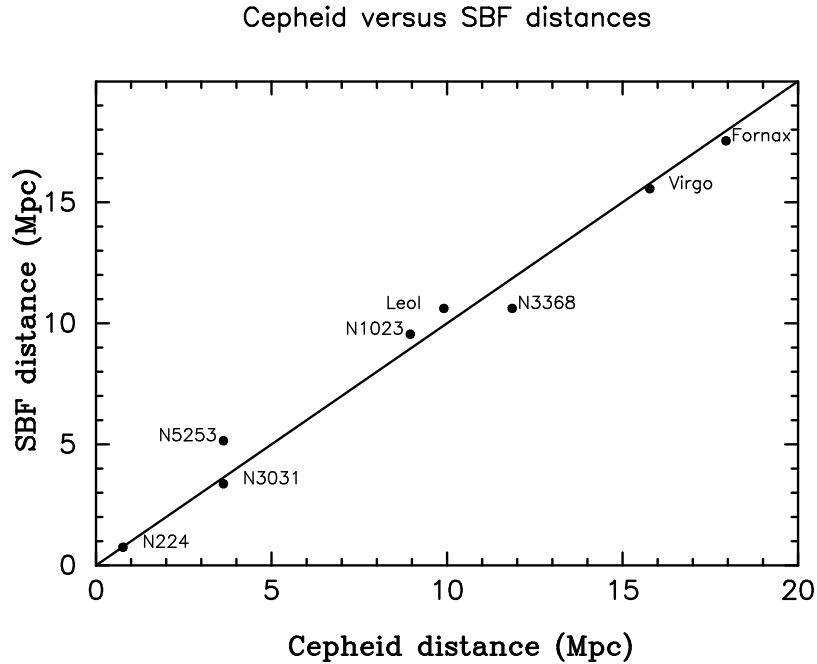}{4.0in}{270}{50}{50}{-195}{295}
\medskip
\caption{ A comparison of distances obtained using Cepheids and the
surface-brightness-fluctuation method. The latter data are from Tonry
(1996, private communication).}

\end{figure} 
\medskip

In Figure 3, a comparison of  distances obtained from Cepheids and the
planetary nebula luminosity function is shown from Jacoby (1997).  The
zero point of  the PNLF  distances is fixed   by adopting the  Cepheid
distance modulus to M31 from Freedman  \& Madore (1990);  however, all
subsequent relative  distances for the  other  galaxies are completely
independent.  The relative $rms$ scatter amounts to  less than $\pm$ 
8 $\%$ in distance.

\begin{figure} 
\plotfiddle{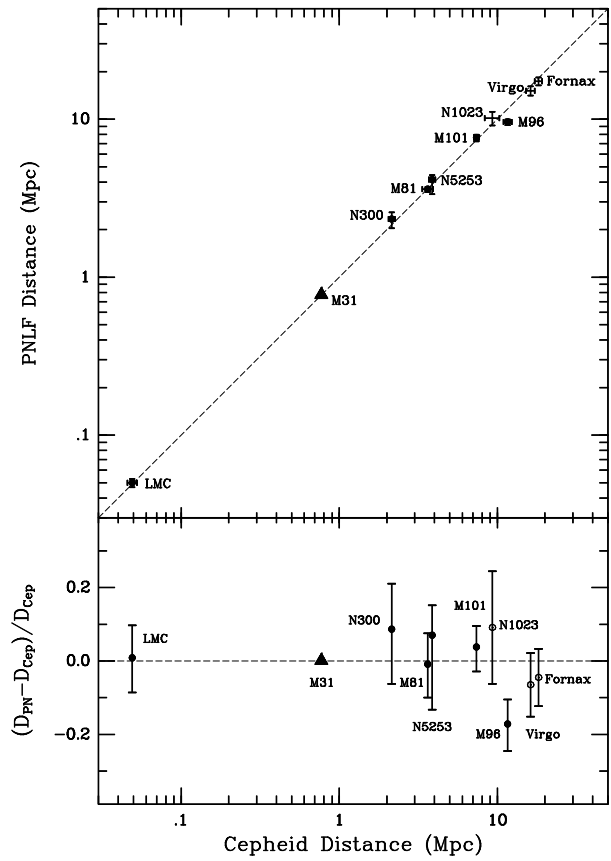}{5.0in}{0}{50}{50}{-160}{-50} 
\medskip
\medskip
\caption{ A comparison of distance moduli obtained from Cepheids compared
with those obtained using  the planetary  nebula  luminosity  function
(PNLF).}
\end{figure} 
\medskip

Madore, Freedman \& Sakai (1997) compare the Cepheid distance scale to
the  RR-Lyrae calibrated tip of  the  red giant branch (TRGB)  method.
Again,  the   results of this  comparison  are noteworthy,  with the
agreement at a level of $\pm$5\% in distance as shown in Figure 4.

The excellent   agreement  in {\it  relative} distances   is extremely
encouraging.    Moreover,  the factors-of-two   discrepancies   in the
distances  to  nearby galaxies have  now been eliminated.  However, it
must be emphasized that there are still disagreements in the {\it zero
points}  of the Cepheid and RR   Lyrae  distance scale  at a  level of
0.15-0.3 mag (8 -  15$\%$ in distance).   For example, as discussed by
Freedman \& Madore (1993), although the Cepheid and RR Lyrae distances
agree  to  within their stated errors,  the differences are systematic
(in the sense that the RR Lyrae distances are smaller than the Cepheid
distances).  This effect has been discussed in detail by Walker (1992)
in the case of the LMC, and Saha {\it et al.} (1992) in the case of IC
1613.  Most recently, this effect has been discussed  by van den Bergh
(1995).  As yet  unresolved  are the slope  and zero   points  of  the
relation between absolute magnitude and  the metallicity  for RR Lyrae
stars, as well  as  the   metallicity sensitivity of   the Cepheid  PL
relations as a function of wavelength. [Note that if the zero point of
the Cepheid distance scale was adjusted by 0.2-0.3 mag consistent with
the  RR Lyrae  scale, the value  of H$_0$ would be {\bf  increased} by
10-15$\%$.]  However, the fact that the Population I Cepheid distances
now agree as well as they do  (to within 0.15  to  0.30  mag) with the
Population II RR  Lyrae and TRGB  distances, is again  consistent with
the predictions of a shallow  metallicity dependence of the Cepheid PL
relations ( Chiosi {\it et al.} 1993), and the observational result of
Freedman \& Madore (1990).

\begin{figure} 
\plotfiddle{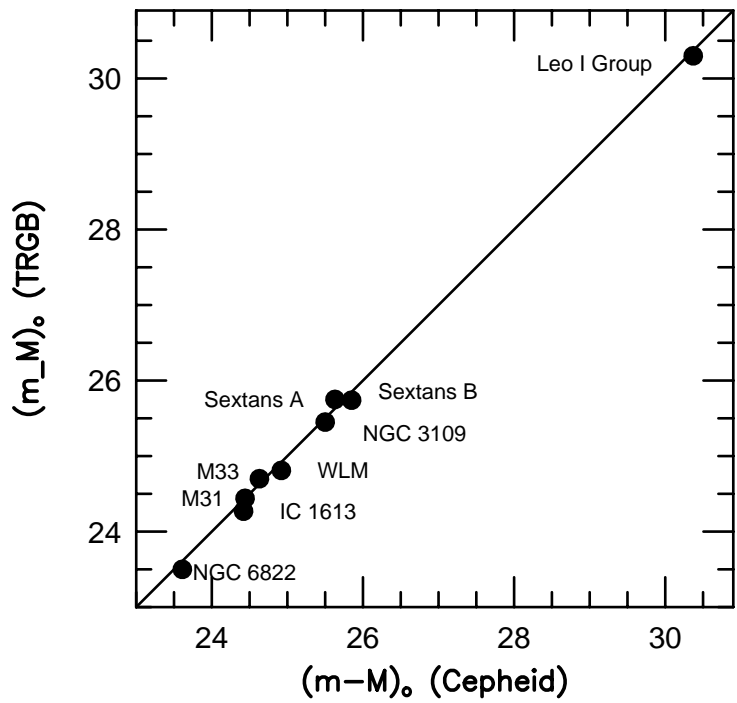}{3.5in}{0}{50}{50}{-160}{-50} 

\medskip
\caption{ A comparison of distances obtained using Cepheids and the
tip  of the red  giant  branch method (Sakai {\it et al.} 1997).}

\end{figure} 
\medskip

It   should  be noted   that   Dr.  Tammann completely  dismissed  the
excellent agreement in these distance  indicators as being unworthy of
discussion.  However,   few of the   distance   indicators adopted  by
Sandage   \&   Tammann  (for   example,  novae, the   globular cluster
luminosity function, or   angular diameters of spiral  galaxies)  have
been  tested with  the same rigor  that has been applied to   the SBF,
PNLF, or TRGB methods.  The  latter  methods have been  compared  on a
case-by-case basis with Cepheid distances, investigated for extinction
and  metallicity  effects,  and  their   relative agreement has   been
quantified, and moreover, found to be excellent.  Before we get to the
bottom of the H$_0$ debate, {\it all} distance indicators will have to
be scrutinized to the same degree.  If  there are systematic errors in
individual distance    indicators,  more  of  this   kind of  detailed
comparison of independent methods will be needed to reveal them.

\section{ Measuring H$_0$ to 10$\%$}


The 1980's and 90's have spawned a wealth of  new efforts in measuring
accurate  relative distances to galaxies    using  a broad   range  of
different techniques.   For  the first  time   in the history  of this
difficult field, relative distances to galaxies can be compared  on an
individual   case-by-case basis,   and their quantitative agreement is
being established.  The reader is  again referred to  the recent STScI
conference proceedings  on the  Extragalactic Distance Scale  for more
detailed  discussions  of these methods (Livio  \& Donahue 1997).  The
discussion  in this section parallels very   closely,  and  provides a
summary of the  results, presented in  that volume by Freedman, Madore
and  Kennicutt (1997).   The Baltimore and  Princeton conferences took
place within  a  month of   each other; hence   no  newer results  are
presented here.

Cepheid distances lie at  the heart of  the Hubble Space Telescope Key
Project on  the Extragalactic  Distance  Scale (Freedman {\it et  al.}
1994a,b; Kennicutt, Freedman \& Mould 1995); and Cepheids are employed
in several other HST distance scale programs ({\it e.g.}, Sandage {\it
et al.} 1996;  Saha {\it et  al.} 1994, 1995; and  Tanvir {\it et al.}
1995).

The  HST Key Project   on  H$_0$   has been  designed to  use  Cepheid
variables to  determine    (Population~I)   primary   distances  to  a
representative sample of galaxies  in the  field, in small groups, and
in major clusters.  The  galaxies were chosen  so that  each   of  the
secondary distance  indicators with measured high internal  precisions
can be accurately calibrated in  zero point, and then intercompared on
an   absolute basis.  The  Cepheid    distances  can then  be used for
secondary calibrations and applied  to independent  galaxy  samples at
cosmologically significant distances.  Cepheid  distances to the Virgo
and  Fornax clusters provide  a   consistency  check  of the secondary
calibrations.  The aim is to derive a  value for the expansion rate of
the Universe, the Hubble constant, to an accuracy of 10$\%$.

A  measurement  of the  Hubble   constant to 10$\%$  accuracy poses an
immense  challenge given the   history  of systematic  errors  in  the
extragalactic  distance scale.  For  this  reason, the Key Project has
been designed to incorporate many independent cross-checks of both the
primary and secondary distance  scales.  The Key Project  is described
in more detail in Kennicutt,  Freedman  \& Mould (1995), Freedman {\it
et al.} (1994a), and  Mould {\it et  al.} (1995).  Briefly,  there are
three  primary goals:  (1)  To discover  Cepheids, and thereby measure
accurate distances to spiral galaxies suitable  for the calibration of
several independent secondary  methods.  (2)  To   make direct Cepheid
measurements of distances  to  three  spiral galaxies  in  each of the
Virgo and   Fornax clusters.  (3)  To  provide  a check   on potential
systematic errors both in the Cepheid distance scale and the secondary
methods.

\subsection{Summary of Recent Results}

All three aspects of  the Key Project  are now well  underway.  Midway
through our   three-year  program,  we have  discovered  over 500 new
Cepheid variables in 9 galaxies.   Most of these  galaxies were chosen
to provide distances critical to the calibration of secondary distance
methods.  In the  case  of the face-on  spiral  galaxy M101,  we  have
measured samples of Cepheids at two different  radial positions in the
disk,   allowing  us  to begin  undertaking  a test  of  the level  of
sensitivity  of  the Cepheid   period-luminosity relation to  chemical
composition.

Prior  to the refurbishment of  the telescope,  observations were made
using WF/PC1 in two fields in the nearby galaxy M81, in  addition to a
field in the outer regions  of M101.  Since the refurbishment  mission
in  December  of  1993,   the    pace  of the  program  has  increased
considerably.  Data have been acquired for several  galaxies: M101 (an
inner  field), the Virgo cluster  galaxy  M100;  eight inclined spiral
galaxies: NGC~925  (a  member  of   the NGC~1023   group),   NGC~7331,
NGC~3621, NGC~2541, NGC~2090, NGC~3351 (a member of  the Leo I Group),
NGC~4414  (a  host galaxy for  a type  Ia  supernova),  and NGC  3198.
Recently, data have also been acquired for NGC~1365, a barred spiral galaxy
located  in the southern  hemisphere  cluster  Fornax,  and   for  two
additional galaxies in the Virgo cluster, NGC 4535 and NGC 4548.

To date  the H$_0$  Key  Project  results have  been  published for  6
galaxies in  the following papers: M81  (Freedman {\it et al.} 1994b),
M100 (Freedman {\it et  al.}   1994a; Ferrarese {\it  et  al.}  1996);
M101 (Kelson  {\it et al.}   1996),  NGC~925 (Silbermann {\it  et al.}
1996), NGC~3621 (Rawson {\it et al.} 1997),  and NGC~3351 (Graham {\it
et al.}   1997).  These distances  are listed in  Table  1, which also
includes our   new, unpublished distance   to NGC~1365.   As  reported
by Dr. Tammann at this conference,  significant progress has  also been made in
the HST supernova calibration  project: Cepheids have been located and
studied in IC~4182 (Saha  {\it et al.}  1994),  NGC~5253 (Saha {\it et
al.} 1995) and NGC~4536 (Saha {\it et al.}  1996).  First results have
also  been published for NGC~4639  and NGC~4496A (Sandage {\it et al.}
1996).  Cepheids have been detected in the Leo I galaxy NGC~3368 (M96)
by Tanvir {\it et al.} (1995).

\medskip

\begin{table} 
  \begin{center} 
  \caption{ KEY PROJECT CEPHEID DISTANCES TO DATE }
  \begin{tabular}{ccr} 
         &          &  \\
  Galaxy & $\mu _0$ & d (Mpc)  \\[3pt]
 NGC 3031    & 27.80$\pm$0.20    & 3.6$\pm$0.3   \\
 NGC 5457    & 29.34$\pm$0.17    & 7.4$\pm$0.6   \\
 NGC 0925    & 29.84$\pm$0.16    & 9.3$\pm$0.7   \\
 NGC 3351    & 30.01$\pm$0.19    & 10.1$\pm$0.9   \\
 NGC 3621    & 29.17$\pm$0.18    & 6.8$\pm$0.7   \\
 NGC 4321    & 31.04$\pm$0.21    & 16.1$\pm$1.5   \\
 NGC 1365    & 31.32$\pm$0.19    & 18.4$\pm$1.6   \\
  \end{tabular}
  \end{center} 
\end{table} 
\smallskip

The current  limit for measuring Cepheid distances  with  HST is about
3,000 km/sec, (Zepf {\it et al.,} 1997),  a distance at which peculiar
velocities can  still be  a substantial  ($>$10$\%$) component of  the
overall cosmic expansion velocity.  To date, the  most distant objects
observed as part of the Key Project have velocities of less than 1,500
km/sec and  require  very  large amounts  of telescope   time (over 30
orbits of  HST time per galaxy).   Hence,  the measurement of H$_0$ to
$\pm$10\% cannot be achieved with  Cepheids alone and  the main thrust
of  the  Key Project  remains   the calibration of secondary  distance
indicators.  However, direct Cepheid distances to the Virgo and Fornax
clusters   with  velocities $\geq$1,200  km/sec  can  still  provide a
consistency check at a level of $\pm$20\%.

A  preliminary estimate of the  Hubble  constant was given by Freedman
{\it et al.} (1994a) and  by Mould {\it et  al.}  (1995), based on the
distance   to M100 in  the  Virgo  cluster.   Initially  a search  for
variables was  conducted using only a  subset of the Wide-Field camera
chips, and twenty high signal-to-noise Cepheids, having periods in the
range of 20 to 65 days, were identified (Freedman {\it et al.} 1994a).
A continuing analysis eventually yielded a larger total sample of over
50  Cepheids, and additional calibration  data were  used to provide a
refined zero point yielding a reddening-corrected  distance to M100 of
16.1 $\pm$1.3 Mpc (Ferrarese {\it et al.} 1996).

As discussed in detail in  Freedman {\it et  al.} (1994a), one  of the
dominant uncertainties in the  determination   of H$_0$ based on   the
Virgo  cluster  is due  to  the  fact that the distribution  of spiral
galaxies  is both extended and  complex.  Hence, the distance  to M100
alone  cannot  define the mean   distance to the   Virgo cluster to an
accuracy of better than 15--20$\%$ (Freedman {\it et al.} 1994a, Mould
{\it et  al.} 1995).  Adopting  a   recession  velocity for  the Virgo
cluster of 1,404 $\pm$80 km/sec (Huchra  1988) and a Virgo distance of
16.1 Mpc (Ferrarese {\it et  al.}  1996) yields a  value of H$_0$ = 87
$\pm$6   (random) $\pm$16  (systematic) km/sec/Mpc.     Alternatively,
adopting a  recession  velocity  of  1,179 $\pm$17 km/sec  (Jerjen and
Tammann 1993) results in  H$_0$ =  73 $\pm$14  km/sec/Mpc for the same
distance.  The  dominant sources of  uncertainty in this estimate  are
systematic  and are due to (a)  the reddening correction, (b) the zero
point of  the  Cepheid PL  relation,  (c)  the position  of M100  with
respect to the center of  the  cluster, and (d) the  adopted recession
velocity of the cluster. Cepheid distances  to 5 galaxies in the Virgo
cluster have now been published, and they  are listed  in Table 2. Data
for two additional Virgo cluster galaxies, NGC 4548 and  NGC 4535, are
currently being analyzed as part of the Key Project.

\medskip

\begin{table} 
  \begin{center} 
  \caption{ CEPHEID DISTANCES TO VIRGO CLUSTER GALAXIES }
  \begin{tabular}{lcc} 
         &                  &           \\
  Galaxy & Distance Modulus & Distance (Mpc) \\[3pt]
 NGC 4321  & 31.04 $\pm$ 0.21 & 16.1 $\pm$ 1.5  \\
 NGC 4496A  & 31.13 $\pm$ 0.10 & 16.8 $\pm$ 0.8  \\
 NGC 4571  & 30.87  $\pm$ 0.15 & 14.9 $\pm$ 1.2  \\
 NGC 4536$^1$  & 31.10 $\pm$ 0.13 & 16.6 $\pm$ 1.0  \\
 NGC 4639  & 32.00  $\pm$ 0.23 & 25.1 $\pm$ 2.5  \\
    &    &    \\
 Mean  & 31.25  $\pm$ 0.45 & 17.8 $\pm$ 4.1    \\
    &    &    \\
  \end{tabular}

$^1$ For consistency, N4536 is corrected for the ``long" zero point by +0.05 mag

  \end{center} 
\end{table}

\medskip

One way to avoid the error due to the uncertainty in the Virgo cluster
velocity is to tie into more  remote clusters (using relative distance
indicators) and step out to a distance where peculiar velocities are a
smaller fractional contribution  to the  overall  expansion  velocity.
For example, an  estimate of H$_0$ can   then  also be made using  the
measured  relative distance  between  the Virgo  cluster  and the more
distant  Coma cluster (Freedman   {\it et al.}   1994a).   Adopting  a
distance  of 16.1 Mpc  for the Virgo  cluster, a Coma distance of 88.7
Mpc (based on a relative Virgo-Coma distance modulus of 3.71 mag), and
a recession velocity for Coma of 7,200 km/sec, yields a value of H$_0$
= 81 $\pm$6  (random) $\pm$15 (systematic) km/sec/Mpc.  These  results
indicate that the value of the Hubble  constant is $\sim$80 km/sec/Mpc
out to a  distance  of ~100 Mpc,  with  an estimated  uncertainty   of
$\pm$20$\%$.

Coma is but one example of a cluster for which relative distances have
been measured  using  a  variety of secondary  methods.   As described
above, we  have   now measured Cepheid   distances  to  a total of   8
galaxies, in addition to M100 in Virgo, most recently, NGC~1365 in the
Fornax cluster.  A preliminary value of the Hubble constant based on a
calibration   of the Tully-Fisher  relation for  a  sample of  distant
clusters is given by Mould {\it et al.} (1997) and also by Madore {\it
et al.}  (1997).   Based on  our new Cepheid   distance to the  Fornax
cluster,  a value of H$_0$  is determined by   tying  into the distant
cluster frame  defined by Jerjen and  Tammann  (1993).  Furthermore, a
Cepheid  distance  to  the   Fornax  cluster  provides  two additional
calibrators for the type Ia supernova distance scale. These recent results
are summarized below.

\subsection{New Results on the Distance to the Fornax Cluster}

Before  proceeding any  further, I   note that  during his debate, Dr.
Tammann referred to the  Fornax cluster as ``the   worst place  in the
Universe  in  which  to calibrate  the Hubble constant''. [Instead, he
argued, the  Virgo cluster is much  more suitable and  he proceeded to
determine the  Hubble constant  using 6 different methods tied  to the
Virgo  cluster.]  Beyond  the   hyperbole, there is  a  valid concern:
namely that it is  critical  to  establish that the location of spiral
galaxies  in  the  Fornax cluster (where  Cepheids  can be  found)  is
representative of the elliptical galaxies  in  the cluster (where many
of  the  secondary methods  can    be applied, for  example   type  Ia
supernovae, surface brightness fluctuations and  the  planetary nebula
luminosity function).   However, the case  that Dr.  Tammann presented
is misleading.  He suggested that the spirals in  this cluster are all
fully detached from  the elliptical core  and systematically closer by
about 0.8 mag.  However, these claims are simply  not supported by the
data.

First, there   is no evidence  for a  peculiar spatial segregation  of
spirals and ellipticals in  Fornax as shown in Figure  5.  This figure
shows the spatial distributions of  the  spiral and elliptical members
within 5  degrees of  core the  Fornax cluster having  measured radial
velocities.  As  is common  in clusters, the  spiral  galaxies tend to
avoid the core and to be located at greater radial distances  from the
center  than the  ellipticals.    Furthermore, using the   most recent
compilation   of published relative distance  moduli,  prepared by Dr.
Tammann's own  doctoral  candidate   (Schroder 1996),  the  unweighted
global  average   of  14 methods is   $\Delta \mu_{F-V}$ =  -0.06 mag.
Excluding the type Ia supernovae, Population I (spiral) moduli average
to -0.20 mag; and Population  II  (elliptical) moduli give -0.08  mag.
There is little  statistical  significance to  that difference of 0.12
mag which, in any case,  amounts to  only 6\% (or 1  Mpc) in distance.
The SNIa modulus stands out  at +0.36 mag,  more  than 0.4 mag fainter
than  the  mean  of  all   other  estimates.  Phillips  (1996, private
communication) stresses  that many of  the  observed supernovae in the
Virgo  cluster were  observed with older, photographic photometry, and
should be discarded  from this type   of analysis.   [Fortunately, the
differential (Fornax minus Virgo) distance moduli are not required for
any further analysis, given a distance modulus  to  the Fornax cluster
based directly on Cepheids.]

\begin{figure} 
\plotfiddle{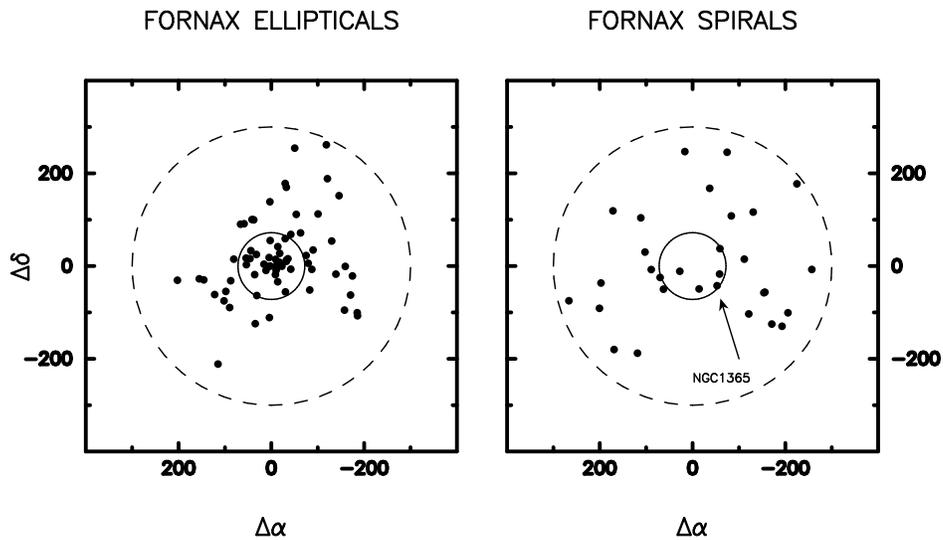}{3.0in}{270}{50}{50}{-195}{295}
\medskip
\caption{ The spatial distribution of 69 elliptical and 30 spiral
galaxies  in the  Fornax   cluster.  Units are   in  arcminutes. These
galaxies  lie  within  5  degrees of  the  Fornax cluster   core, have
published radial velocities less than 3,000 km/sec,  and were found in
a search of  the  NED database.  Left Panel:   E/SO   galaxies.  Right
Panel:  Spiral  and  irregular galaxies.   The  outer  broken  circles
correspond   to the sample  search   radius.  The inner  solid circles
correspond to the  radial distance of NGC  1365 from the center of the
Fornax cluster and  are roughly equivalent to  the core radius  of the
cluster as a whole. }
\end{figure} 
\medskip

Second,  based on an  analysis of  redshifts  for  99 Fornax galaxies,
there is also no observational evidence for a  velocity segregation of
the spiral and elliptical populations: the mean radial velocity of the
30  member spirals and irregulars with  published radial velocities is
+1,476$\pm$335 km/sec,  and the mean   for 69 ellipticals and S0's  is
+1,451$\pm$311   km/sec.  Again,     the difference (25   km/sec)   is
statistically  insignificant, given that  the velocity dispersions  of
the two populations are both in  excess  of $\pm$300 km/sec,  and that
neither mean is determined to better than that calculated difference.

Quite the contrary  to Dr.  Tammann's  remarks, as  shown  below,  the
Fornax cluster appears to  provide an excellent opportunity to compare
and contrast different secondary techniques.  The  Fornax cluster is a
particularly  important   cluster for a  number  of   reasons.  First,
because it is  a very compact cluster (see  Figure  6), it provides  a
calibration of several secondary methods.  Of particular importance is
that it contains two well-observed recent type Ia supernovae, allowing
a  direct comparison between  the   type Ia distance   scale and other
well-studied secondary indicators with small measured dispersions such
as  the Tully-Fisher   relation, surface  brightness fluctuations, and
planetary nebula luminosity function.

There are several independent  pieces of  evidence that NGC~1365  is a
{\it  bona-fide}  member of the  Fornax cluster  (Madore {\it et  al.}
1997).  First, NGC~1365 is an optical member of the cluster as it lies
directly along  our line of sight to  Fornax, projected only  $\sim$70
arcmin  from the geometric center  of the  ($\sim$200~arcmin diameter)
cluster  itself.  Second, NGC~1365  is also coincident with the Fornax
cluster in velocity space; its velocity off-set from  the mean is only
half   of  the   overall  cluster velocity  dispersion.   Hence,  both
positional and velocity  constraints place NGC~1365 physically  in the
Fornax cluster.  Finally, NGC~1365 sits only 0.02~mag from the central
ridge  line of the   {\it  apparent}  Tully-Fisher relation for  other
cluster members defined by recent studies of the  Fornax cluster ({\it
e.g.,} Bureau, Mould  \& Staveley-Smith 1996,  Schroder  1996).  (Note
that NGC~1365 is  amongst the  most  luminous nearby galaxies   on the
basis of the Tully-Fisher  relation.  It is brighter than  M31 or M81,
but comparable  to NGC~4501 in  the Virgo cluster   or NGC~3992 in the
Ursa Major cluster.)

At present we are analyzing a  sample of 53  newly discovered Cepheids
in the galaxy  NGC~1365 in Fornax.   Details of the  observations  for
NGC~1365 will  be given by Madore {\it  et al.}  (1997) and Silbermann
{\it et al.} (1997).  A preliminary true distance modulus of $\mu_0 =$
31.32  $\pm$0.12~mag  is obtained  based on  V and I period-luminosity
relations.   This  corresponds   to   a distance  to NGC~1365  of 18.4
$\pm$1.0~Mpc.

\subsection{ Estimates of H$_0$ Based on the Distance to the Fornax Cluster}

Three different estimates of H$_0$ based on the distance to the Fornax
cluster are presented here;  the reader is  referred to Madore {\it et
al.} (1997) for more  details.  The first estimate is  based solely on
the velocity  and the  Cepheid distance  to  the Fornax cluster.   The
second  estimate  is based on  the nearby volume of  space,  up to and
including both the  Virgo and  Fornax clusters.   The  third  estimate
comes from using the Cepheid distance to Fornax to lock into secondary
distance indicators, thereby allowing us to step out to cosmologically
significant  velocities (10,000~km/sec and  beyond)  corresponding  to
distances on the order of 100~Mpc. The first two estimates are subject
to larger uncertainties due to the local flow field.

\subsubsection{ The Hubble Constant at Fornax\ \ \  \ } 

As described earlier, one of the largest uncertainties in the estimation
of H$_0$ based on the distance to  M100 in the Virgo cluster (Freedman
{\it et  al.} (1994a), is  the issue of   the line-of-sight positional
uncertainty of M100  relative   to   the elliptical-rich  core  of the
cluster. The  second  major  uncertainty  is  due  to   the  uncertain
Virgo-centric     flow  velocity   correction   for the   Local  Group.
Fortunately, the  situation   for the  Fornax   cluster  is far   less
uncertain.

Figure 6 shows a comparison of the two clusters  of galaxies  drawn to
scale,  as seen  projected on the sky.  Both  clusters are  located at
very similar distances from  us.   However, as can  be clearly seen in
this figure, NGC~1365 is projected significantly closer to the core of
the Fornax  cluster than is  M100 with  respect to the  ellipticals in
Virgo.  Furthermore, since  the Fornax  cluster is considerably   more
centrally  concentrated  than   Virgo,  the  back-to-front uncertainty
associated with its  three-dimensional spatial  extent is dramatically
reduced for any randomly selected member.  Both the compactness of the
Fornax  cluster and the  actual proximity  of NGC~1365 to  the core of
that cluster result  in a  statistical uncertainty of only  2-3\% when
taking  the Cepheid  distance to NGC~1365  and identifying it with the
distance to the core of the Fornax cluster group (Madore {\it et al.,}
1997).

\begin{figure} 
\plotfiddle{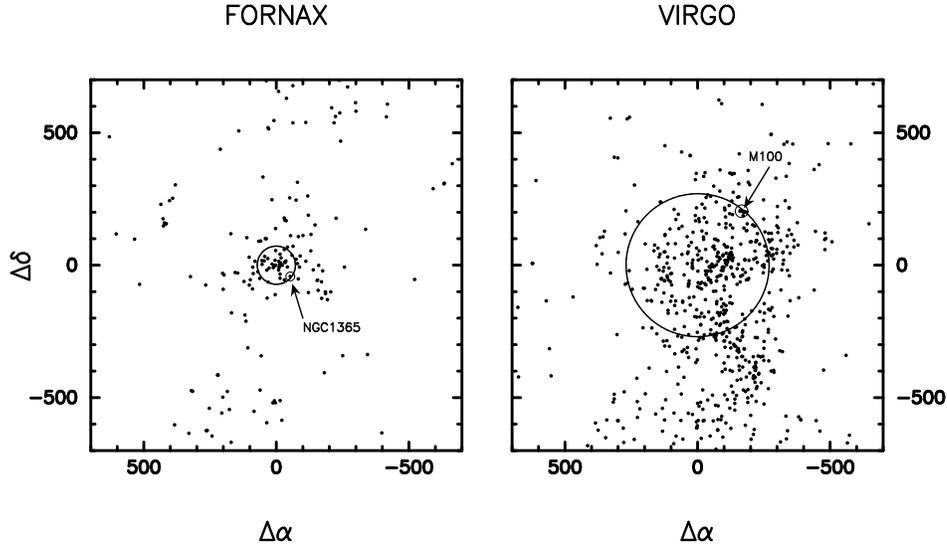}{3.0in}{270}{50}{50}{-195}{290}
\medskip
\medskip
\caption{ The  distribution of galaxies  as projected on  the  sky for
the Virgo  cluster (right panel) and  the Fornax cluster (left panel).
The positions of M100 and NGC~1365 are marked by arrows. 
The units are arcminutes. }
\end{figure} 
\medskip

Adopting our Cepheid distance of 18.4~Mpc  as representative of Fornax
cluster gives  H$_0=$72  $\pm$3~~$\pm$18  km/sec/Mpc, for an   adopted
cosmological  expansion rate of  1,300~km/sec.   The first uncertainty
includes random errors in the distance from the PL fit to  the Cepheid
data,  as well as  random velocity errors  in the adopted Virgo-centric
flow, combined with the distance   uncertainties to Virgo   propagated
through  the  flow model.   The   second uncertainty represents    the
systematic errors associated with the adopted mean velocity of Fornax,
and the adopted zero point of the PL relation (combining in quadrature
the LMC distance   uncertainty  and   a  measure  of  the  metallicity
uncertainty).

{\subsubsection{  The Nearby Flow Field \ \ \ \ \  \ \ \ } 

In  Figure 7,  a Hubble diagram  (in the  sense originally  plotted by
Hubble   (1929))  of  distance     versus  velocity   is   shown.  The
galaxies/groups in this  plot all  have distances determined  directly
from Cepheids.   The  expansion velocities  are individually corrected
for Virgo-centric flow    (using  a Local  Group  infall  velocity   of
200~km/sec.)  At 3~Mpc the M81-NGC~2403 Group (for which both galaxies
of this  pair have Cepheid   distance determinations)  gives H$_0   =$
75~km/sec/Mpc.  Working further  out  to M101, the NGC~1023  Group and
the  Leo Group, the calculated  values  of  H$_0$ vary from  65  to 95
km/sec/Mpc.   An   average  of    the  six  independent determinations
including Virgo and Fornax, gives H$_0$ = 75 $\pm$8~km/sec.

\begin{figure} 
\plotfiddle{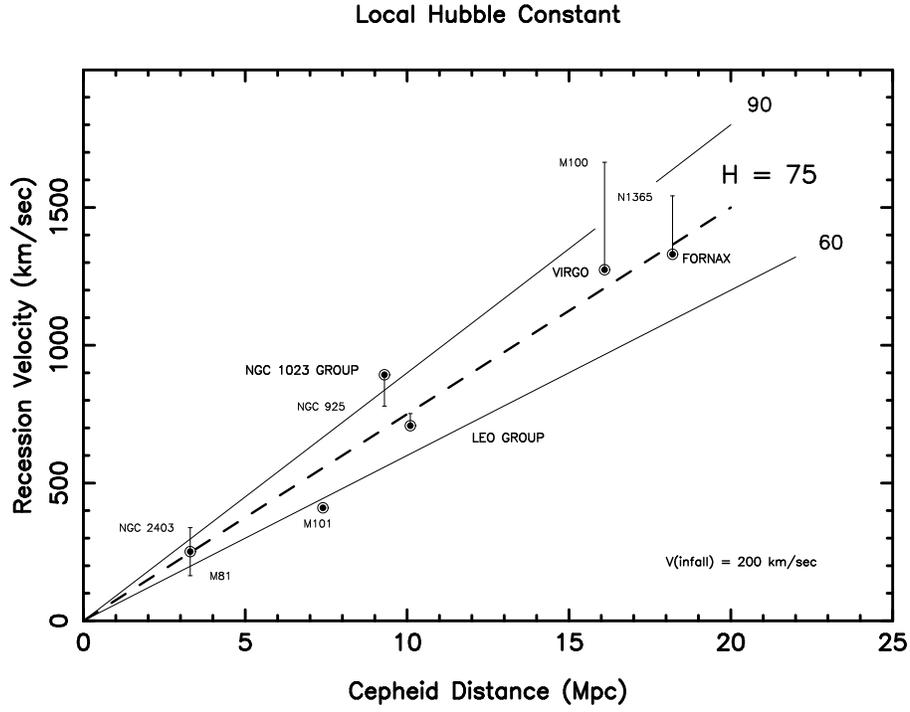}{3.5in}{270}{50}{50}{-195}{295}
\medskip
\medskip
\caption{ The  Hubble velocity-distance  relation for nearby
galaxies having Cepheid distances.  Circled dots denote the velocities
and  distances of the parent  groups or clusters.  The velocities  for
the individual  galaxies are also indicated at  the ends of the  bars.
The broken  line  represents a  fit  to   the data  giving H$_0   =  $
75 $\pm$ 15 (solid lines) ~km/sec/Mpc.}
\end{figure} 
\medskip

Although at these nearby distances one expects a large scatter due to
peculiar velocities, an interesting feature  about this figure is  how
remarkably low in dispersion this  Hubble relation is.  More distances
are  required  to  ascertain  whether this  low  scatter  is  simply a
consequence of  small-number statistics, or whether  it is signaling a
true, quiet, local flow  (beyond the influence of the Virgo cluster).

The two determinations of H$_0$ in  this and in the preceding section
make    no explicit    allowance    for  the possibility  that     the
inflow-corrected velocities of  both the  Fornax  and  Virgo  clusters
could  be  perturbed  significantly by  other mass   concentrations or
large-scale  flows.   However, it is  interesting  to note  that these
local estimates do agree very well with the determinations of H$_0$ at
larger distances, where peculiar velocities are a fractionally-smaller
uncertainty.

\medskip
\medskip
\medskip
\medskip

{\subsection{  H$_0$ via Secondary Indicators} \ \ \ \ \  \ \ \ } 

Although  these preliminary estimates of  H$_0$  from Cepheids  in the
Virgo (M100) and  Fornax  (NGC~1365) clusters  are  illuminating,  the
primary route to H$_0$ in the Key Project is  through  the calibration
of secondary indicators, which    can extend the distance  scale  well
outside the local supercluster.  To avoid these local uncertainties we
step out from Fornax to the distant flow field based on: (1) using the
distance  to Fornax  to  tie into   averages over previously published
differential moduli for independently selected distant-field clusters;
(2) recalibrating the type Ia supernova luminosities at maximum light
and applying that  calibration  to events as  distant as 30,000~km/sec;
and (3)  using  these data   to calibrate the  Tully-Fisher
relation ( Mould {\it et al.,} 1997).

\subsubsection{  Beyond Fornax: Distant Clusters }

Jerjen \&  Tammann (1993) have   compiled a set  of  relative distance
moduli  based  on   their evaluation  and   averaging of  a  number of
independent secondary distance indicators, including brightest cluster
galaxies, the Tully-Fisher  relation,  and supernovae.   They  conclude
that this sample is   ``minimally biased''.  We have  adopted  without
modification  their  differential  distance scale  and  tied into  the
Cepheid distance  to the   Fornax   cluster, which was  part of  their
sample.  The  results are  shown  in  Figure  8   which extends   the
velocity-distance relation out to more than 150~Mpc. No error bars are
given  in the   published compilation, but   it is clear from the  plot
itself that  the observed scatter  can be fully  accounted for by 10\%
errors  in  distance and/or   velocity.  This  sample  is sufficiently
distant to average over the potentially biasing effects of large-scale
flows, and yields a value of H$_0 =$ 72 $\pm$4~km/sec (random), with a
systematic error  of 10\% being  associated with the distance (but not
the velocity) of the Fornax cluster.   Again, the coincidence of H$_0$
measured at Fornax with that for the far field seems to indicate that
Fornax itself  does not have a large  component of motion with respect
to the microwave background.

\begin{figure} 
\plotfiddle{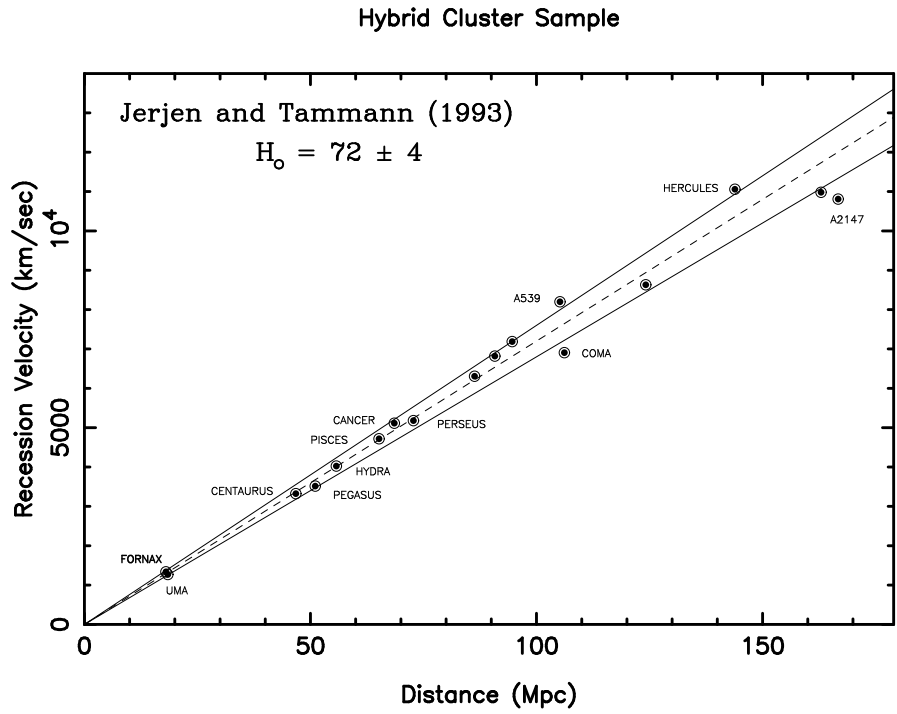}{3.0in}{270}{50}{50}{-195}{295}
\medskip
\medskip
\caption{  Recession velocity  versus  distance  for  a sample  of  17
distant clusters from the published data of Jerjen  \& Tammann (1993).
Absolute  distances are  calculated  by  tying   the  relative cluster
distances to the Cepheid distance of the Fornax cluster.}
\end{figure} 
\medskip

\subsubsection{ Calibration of  Type Ia Supernovae \ \ \ \ \ \  }

The Fornax cluster elliptical galaxies  NGC~1316 and NGC~1380 are host
to the well-observed type Ia supernovae 1980N and 1992A, respectively.
(The supernova  1981D was also observed in  NGC~1316, but the data are
photographic, and hence are not  of as high  quality as the other two,
and  will not be   considered  here.)  The  new  Cepheid  distance  to
NGC~1365,  and associated   estimate of  the  distance  to  the Fornax
cluster discussed above,  thus  allow two additional very  high-quality
objects to be added to the calibration of type Ia supernovae.  Details
of this  calibration are  given in Freedman   {\it et al.}   (1997, in
preparation) where    corrections   for  interstellar  extinction  and
decline-rate correlations  are presented.  Application  to the distant
Type Ia supernovae of   Hamuy (1995) gives H$_0$  = 63--68~km/sec/Mpc.
The range of values  reflects different choices  of weighting to  both
the calibrator galaxies, in addition to the distant supernova sample.

In Figure 9, the absolute-magnitude decline-rate  relation for type Ia
supernovae having direct Cepheid distances  (top panel)  is shown; for
comparison the  lower panel shows  the same relation  for  the distant
supernova sample of Hamuy {\it et al.} (1995) and Phillips (1993).  It
should  be   recalled that  in  the cases of   1980N  and  1992A,  the
supernovae occurred in elliptical galaxies in the Fornax cluster (that
is, not in NGC~1365 for which the Cepheid distance has been measured).
The same is true of SN 1989B (Sandage { \it et al.} 1996), although in
this case the association  of  the host galaxy  NGC~3627 with  the Leo
triplet is not  very  well established. 


The Cepheid-calibrating galaxies  provide confirming  evidence  for an
absolute-magnitude decline-rate  relation  for  type Ia  supernovae as
suggested by Phillips (1993), Hamuy {\it et al.} (1995), Reiss, Press,
\& Kirshner (1996), and are contrary to the earlier arguments on this
issue  by  Sandage {\it  et al.} (1996).   The  larger value  of H$_0$
reported  here compared to  that of  Sandage {\it  et al.}  (1996) (57
km/sec/Mpc) is due to three factors:  (1) low weight  is given here to
historical  supernovae observed  photographically, (2)  a decline-rate
absolute-magnitude relation has been included, and (3) the addition of
the  new Fornax calibrators.   All three factors contribute in roughly
comparable proportions.

\begin{figure} 
\plotfiddle{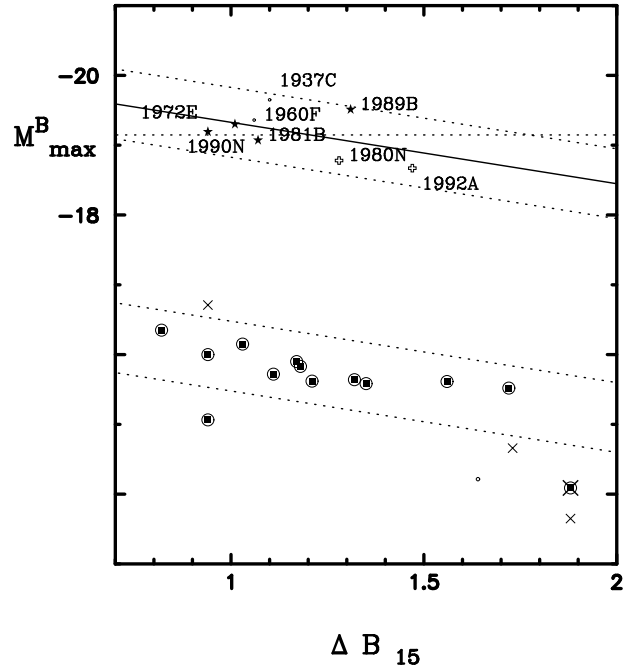}{3.0in}{270}{50}{50}{-195}{290}
\medskip
\caption{ Decline rate as a function of absolute B magnitude. The
points in the top half of the plot are for  type Ia supernovae located
in galaxies   for  which Cepheid distances  have   been  measured.  An
arbitrary magnitude scale is  indicated for the distant  supernovae in
the  bottom part of the  panel.  The X's   in the bottom  panel denote
supernovae with peculiar spectra (SN 1986G,  1991T, and 1991bg) and SN
1992J, a supernova with a red (B-V$>$0.5 mag) color.  The small circle
is for 1971I. The slope of the solid line is determined from the Hamuy
{\it et al.} (1995) sample. }
\end{figure} 
\medskip

\subsection{ Comparison of Cepheid and   Type II Supernova Distances\ \ \ \ \ \  }

\medskip

In Table 3, a list of the  current  galaxies  having both Cepheid and
type II supernova (SNII) distances is  given.   The SNII distances are
from Schmidt, Kirshner,
\&  Eastman (1992)  and  updated by Kirshner  (private communication).
The agreement is excellent: the  mean  ratio of distances amounting to
0.96.

The results of   this and  earlier  comparisons  (see  Section 3.2)  are
striking. The underlying  physics of expanding  supernova atmospheres,
He-flash   red giant    stars, Cepheids, and  planetary  nebulae   are
completely independent.  Yet, the relative distances for these methods
are in  remarkable agreement.  These  results offer encouragement that
the   systematic  errors      that  have traditionally    affected the
extragalactic distance scale are   being very  significantly  reduced.
Moreover, the current  uncertainties  will continue to  be  reduced as
ongoing large ground-based programs, as  well as  the HST  Key Project,
reach completion.

\begin{table} 
  \begin{center} 
  \caption{COMPARISON OF CEPHEIDS AND TYPE II SUPERNOVAE  }
  \begin{tabular}{lcccl} 
        &         &     &             &             \\
  ~Host & Cepheid & SN II & Cepheid/SN II & Supernova   \\
Galaxy& Distance & Distance & Distance Ratio  & ~~~Name   \\[3pt]
 LMC & 0.50 $\pm$ 0.02&0.49 $\pm$ 0.03 & 1.02 $\pm$ 0.08  & SN 1987A  \\
 M81 &  3.6 $\pm$ 0.3 & 4.2 $\pm$ 0.6  & 0.86 $\pm$ 0.18  & SN 1993J  \\
 M101&  7.5 $\pm$ 0.7 & 7.4 $\pm$ 1.2  & 0.97 $\pm$ 0.20  & SN 1970G  \\
 M100& 16.1 $\pm$ 1.8~ &15.0 $\pm$ 4.0~  & 1.07 $\pm$ 0.26  & SN 1974C  \\
 NGC 1058&  9.3$^1$ $\pm$ 0.7 & 10.6$^2$ $\pm$ 1.5  &  0.88 $\pm$ 0.23  & SN 1969L  \\
    &    &   &    &   \\
  \end{tabular}

$^1$ Group membership NGC~1023; Cepheid distance to NGC~925

$^2$ Revised distance (Kirshner, private communication)
  \end{center} 
\end{table} 
\smallskip

\medskip

\subsection{ Summary of the H$_0$ Key Project Results}

At the time of writing, data  have been acquired for 12  out of the 18
galaxies  ultimately comprising the   target  sample for the  HST  Key
Project on the Extragalactic Distance Scale. At this mid-term point in
the  program,  our  results  yield a  value  of  H$_0$  = 73   $\pm$ 6
(statistical) $\pm$ 8  (systematic)  km/sec/Mpc.  The systematic error
takes   into  account  a  number  of factors   including:  the present
uncertainty  in the zero  point   of  the  Cepheid   period-luminosity
relation of $\pm$5\% $rms$ (or equivalently the full-range uncertainty
in the distance   to the  LMC);  the   potential  uncertainty  due  to
metallicity, also in the Cepheid period-luminosity relation at a level
of $\pm$5\%  $rms$, an uncertainty  which allows  for  the possibility
that the locally measured  H$_0$ out to  10,000  km/sec may not be the
global value of H$_0$ of $\pm$7\%; plus an allowance for a scale error
in  the photometry that could  affect all of the  results by $\pm$3\%.
{\bf At the  present time,  the  total uncertainties amount  to  about
$\pm$15\%.   Our   current adopted value for  H$_0$    is 73 $\pm$  10
km/sec/Mpc.} This result is based on a variety of methods, including a
Cepheid calibration of the Tully-Fisher relation,  type Ia supernovae,
a calibration  of distant clusters  tied to Fornax, and direct Cepheid
distances out to $\sim$ 20~Mpc.  In Table 4 the values of  H$_0$ based
on these various methods are summarized.

\medskip

\begin{table} 
  \begin{center} 
  \caption{SUMMARY OF KEY PROJECT RESULTS ON H$_0$}
\smallskip
  \begin{tabular}{lc} 
             &         \\
      Method & H$_0$   \\[3pt] 
       Virgo & 80 $\pm$ 17 \\ 
       Coma via Virgo   & 77 $\pm$ 16 \\
       Fornax & 72 $\pm$ 18 \\ 
       Local  & 75 $\pm$ ~8 \\ 
       JT clusters & 72 $\pm$ ~8 \\
       SNIa   & 67 $\pm$ ~8 \\ 
       TF     & 73 $\pm$ ~7 \\ 
       SNII   & 73 $\pm$ ~7 \\ 
       D$_N -\sigma$ & 73 $\pm$ ~6 \\ 
              &   \\
       Mean   &   73 $\pm$ ~4 \\
              &   \\
     
       {\bf Systematic Errors} & {\bf ~~~$\pm$ 4 ~~~~~$\pm$ 4 ~~~~~$\pm$ 5  ~~~~~$\pm$ 2} \\
              & ~~~(LMC) ~([Fe/H]) ~(global) ~(photometric) \\
              &   \\
  \end{tabular}

  \caption{Current  values  of  H$_0$  for  various  methods.   For each
method, the formal statistical uncertainties are given. The systematic
errors (common to all of these  Cepheid-based calibrations) are listed
at the  end of  the table.   The  dominant uncertainties are    in the
distance to the LMC  and the potential effect  of metallicity  on  the
Cepheid PL  relations, plus  an allowance is  made for the possibility
that  locally the measured value  of H$_0$ may differ from  the global
value.  Also  allowance is made for  a  systematic  scale error in the
photometry which might be affecting all software packages now commonly
in use.  Our  best current weighted mean  value is H$_0$ = 73  $\pm$ 6
(statistical) $\pm$ 8 (systematic).}
\end{center}
\end{table}

\medskip

Currently, we estimate  the total  1-$\sigma ~ rms$ uncertainty in the
value of H$_0$ to  be  approximately  $\pm$15\%.  The  formal internal
uncertainties in the individual secondary methods are small ($<$10\%).
The  dominant overall  error   is still   systematic.  All  of   these
uncertainties  will  be reduced in  the   near   term as  more Cepheid
calibrators  become available  as  a   result  the Key   Project,  the
supernova, and  Leo~I  programs,  and  in  later  cycles  as  infrared
observations of   Cepheids become    available once  the near-infrared
NICMOS camera has been installed on HST.

\section{ Concluding Remarks}

Recent  results  on  the extragalactic  distance scale   are extremely
encouraging.   A   large  number    of  independent  secondary methods
(including the  most recent  type Ia supernova  calibration by Sandage
{\it et al.}  1996) appear to be converging on a  value of H$_0$ in the
range of 60 to 80 km/sec/Mpc.  The  factor-of-two discrepancy in H$_0$
appears to be behind us.  However,  these results again underscore the
importance of reducing remaining errors in the Cepheid distances ({\it
e.g.,} reddening and  metallicity), since at present  the majority of
distance   estimators   are tied to  the   Cepheid  distance scale.  A
1-$\sigma$ error of  $\pm$10\%  on H$_0$ (the aim of  the Key Project)
currently amounts to approximately $\pm$ 7 km/sec/Mpc,  and translates
into a 95\% confidence interval of roughly 55 to 85 km/sec/Mpc.

While this  is   an enormous    improvement over  the    factor-of-two
disagreement of the previous decades,  it is not sufficiently precise,
for example,  to discriminate between   current models of  large scale
structure formation, or to  resolve  definitively the fundamental  age
problem, or the possibility  of a non-zero  value of $\Lambda$. Before
compelling  constraints  can be made   on cosmological  models, it  is
imperative to rule out remaining sources  of systematic error in order
to severely limit the alternative interpretations that  can be made of
the data.  The spectacular  success of HST,  and the fact that a value
of  H$_0$ accurate to   10\% (1-$\sigma$) now  appears quite feasible,
also brings into sharper focus smaller (10-15\%) effects which used to
be   buried in the noise  in   the era of factor-of-two discrepancies.
Fortunately, a  significant improvement will be  possible with the new
infrared capability afforded by  NICMOS.  NICMOS  observations planned
for   Cycle 7  will  reduce  the  remaining  uncertainties  due to both
reddening and metallicity by a factor of 3.

\medskip
\medskip

It is a pleasure to thank the organizers of  this cosmology conference
marking  the 250th  anniversary of  Princeton  University  for such  a
lively and stimulating meeting.  I sincerely thank  the  other members
of the HST H$_0$ Key Project  team, whose dedicated  efforts have made
this Key  Project a reality: R.  Kennicutt,  J.R.  Mould (co-PI's), S.
Faber, L.  Ferrarese, H.  Ford, B.  Gibson,  J.   Graham, J.  Gunn, M.
Han, J.  Hoessel,  J.    Huchra, S.   Hughes, G.   Illingworth,   B.F.
Madore,   R.  Phelps,  A.  Saha,  S.    Sakai, N.  Silbermann, and  P.
Stetson, and graduate students F.  Bresolin, P.   Harding, D.  Kelson,
L.  Macri, D.  Rawson, and A.  Turner.  Special thanks to Barry Madore
and Nancy   Silbermann  for permitting me  to present  results  on the
Fornax cluster in  advance of publication, and to  John  Tonry, George
Jacoby and   Shoko Sakai  for providing  me with  unpublished data and
permission to show them.  The work presented in this paper is based on
observations with the NASA/ESA Hubble Space Telescope, obtained by the
Space Telescope Science  Institute, which is operated  by  AURA,  Inc.
under NASA contract No.  5-26555.  Support for this work  was provided
by NASA through grant GO-2227-87A from STScI.





\end{document}